\documentclass{article}
\usepackage{amsmath}

\numberwithin{equation}{section}
\input{tcilatex}
\begin{document}

\centerline{\large\bf   
RANK-BASED SLOCC CLASSIFICATION FOR ODD $N$ QUBITS}

\vspace*{8pt}
\centerline{Xiangrong Li$^{a}$, Dafa Li$^{b}$}
\vspace*{5pt}

\centerline{$^a$ Department of Mathematics, University of
California, Irvine, CA 92697-3875, USA}

\centerline{$^b$ Dept of mathematical sciences, Tsinghua University,
Beijing 100084 CHINA}

\abstract
We study the entanglement classification under stochastic local
operations and classical communication (SLOCC) for odd $n$-qubit
pure states. For this purpose, we introduce the rank with respect to
qubit $i$ for an odd $n$-qubit state. The ranks with respect to
qubits $1,2,\cdots, n$ give rise to the classification of the space
of odd $n$ qubits into $3^n$ families. 

keywords: Entanglement classification, SLOCC
equations, odd $n$ qubits \vspace*{3pt} 

\vspace*{1pt}
\section{Introduction}          %) A SECTION HEADING
\vspace*{-0.5pt} \noindent
%%%%%%%%%%%%%%%%%%%%%%%%%%%%%%%%
%put the text of the paper here
%%%%%%%%%%%%%%%%%%%%%%%%%%%%%%%%
Quantum entanglement plays a crucial role in quantum computation and
quantum information processing. If two states can be obtained from
each other by means of local operations and classical communication
with nonzero probability (SLOCC), then the two states are said to
have the same kind of entanglement \cite{Bennett} and suited to do
the same tasks of quantum information theory \cite{Dur}.

The complete classification for three qubit pure states has been
achieved \cite{Dur}. While there are six SLOCC equivalence classes
for pure states of three qubits, two of which are genuine
entanglement classes: the $|GHZ\rangle$ class and the $|W\rangle$ class, 
the number of
SLOCC equivalent classes for four or more qubits is infinite. An
important first step in tackling the classification problem for four
or more qubits is to divide the infinite SLOCC classes into a finite
number of families, using some type of criteria to determine which
family an arbitrary state belongs to. Many efforts have been devoted
to the SLOCC entanglement classifications for pure states of four
qubits which result in different finite number of families or
classes, including those based on Lie group theory \cite{Moor}, on
hyperdeterminant \cite{Miyake}, on inductive approach \cite{Lamata},
and on string theory \cite{Borsten}. Polynomial invariants for four
and five qubits \cite{Luque03,Luque06,Osterloh09} as well as for $n$
qubits \cite{LDF07a,LDFJPA} have been discussed, and several
attempts have been made for SLOCC classification via the vanishing
or not of the polynomial invariants \cite{LDFJPA,LDF07b,LDF09,
LDFEPL,Zha,Viehmann}, Recently, entanglement classification for the
symmetric $n$-qubit states has been achieved by introducing two
parameters called the diversity degree and the degeneracy
configuration \cite{Bastin}.

In this paper, we investigate SLOCC classification of odd $n$-qubit pure 
states. To this end, we introduce the rank with respect to qubit $i$ for an odd
$n$-qubit state and establish its invariance under SLOCC. The rank
with respect to qubit $i$ ranges over the values 0, 1, 2, and
therefore gives rise to the classification of the space of odd $n$
qubits into 3 families, as exemplified here. Furthermore, the ranks
with respect to qubits $1, 2, \cdots, n$, permit the partitioning of
the space of the pure states of odd $n\geq 5$ qubits into $3^{n}$
inequivalent families under SLOCC. We also characterize pure
biseparable states and genuinely entangled states in terms of the
ranks.

The paper is organized as follows. In section 2, we introduce the
rank with respect to qubit $i$ for any state of odd $n\geq 3$
qubits. In section 3, we investigate SLOCC classification of odd $n$
qubits. We give the brief discussion in section 4 and the conclusion
in section 5.

\section{Rank of a state with respect to qubit $i$}
\noindent For odd $n$ qubits, let the state $|\psi \rangle
=\sum_{i=0}^{2^{n}-1}a_{i}|i\rangle $, where $|i\rangle $ are basis
states and $a_{i}$ are coefficients. Let the $2\times 2$ matrix
\begin{equation}
M(|\psi \rangle )=\left(
\begin{tabular}{cc}
$P(|\psi \rangle )$ & $T(|\psi \rangle )$ \\
$T(|\psi \rangle )$ & $Q(|\psi \rangle )$%
\end{tabular}%
\right) ,  \label{relation-1}
\end{equation}%
where $T(|\psi \rangle )$, $P(|\psi \rangle )$ and $Q(|\psi \rangle
)$ are three quantities defined on the space of pure states of odd
$n$ qubits:
\begin{eqnarray}
T(|\psi \rangle )
&=&\sum_{i=0}^{2^{n-1}-1}(-1)^{N(i)}a_{i}a_{2^{n}-i-1},
\label{T-1} \\
P(|\psi \rangle )
&=&2\sum_{i=0}^{2^{n-2}-1}(-1)^{N(i)}a_{2i}a_{2^{n-1}-2i-1},  \label{P-1} \\
Q(|\psi \rangle )
&=&2\sum_{i=0}^{2^{n-2}-1}(-1)^{N(i)}a_{2^{n-1}+2i}a_{2^{n}-2i-1}.
\label{Q-1}
\end{eqnarray}%
Here $N(i)$ is the parity of $i$ (i.e. the number of 1's in the
binary representation of $i$). Clearly $M(|\psi \rangle )$ is
symmetric and the rank of $M(|\psi \rangle )$ ranges over the values
0, 1, 2. We refer to the rank of $M(|\psi \rangle )$ as the rank of
the state $|\psi \rangle $ with respect to qubit 1.

As the quantities $T(|\psi \rangle )$, $P(|\psi \rangle )$ and
$Q(|\psi \rangle )$ vary under transpositions $(1,i)$ on qubits 1
and $i$ ($2\leq i\leq n$), so in general does the rank of the state
$|\psi \rangle $ with respect to qubit 1. The variance allows one to
define the rank of a state with respect to qubit $i$ ($2\leq i\leq
n$). For this purpose, we first
let $%
T^{(i)}(|\psi \rangle )$, $P^{(i)}(|\psi \rangle )$ and
$Q^{(i)}(|\psi
\rangle )$\ be obtained from $T(|\psi \rangle )$, $P(|\psi \rangle )$ and $%
Q(|\psi \rangle )$, respectively, under transposition $(1,i)$ on
qubits 1 and $i$, namely
\begin{eqnarray}
T^{(i)}(|\psi \rangle ) &=&(1,i)T(|\psi \rangle ),  \label{T-2} \\
P^{(i)}(|\psi \rangle ) &=&(1,i)P(|\psi \rangle ),  \label{P-2} \\
Q^{(i)}(|\psi \rangle ) &=&(1,i)Q(|\psi \rangle ),  \label{Q-2}
\end{eqnarray}%
for $i=1, 2, \cdots, n$. It is trivial to see that $T^{(1)}(|\psi
\rangle
)=T(|\psi \rangle )$,\ $P^{(1)}(|\psi \rangle )=P(|\psi \rangle )$, and $%
Q^{(1)}(|\psi \rangle )=Q(|\psi \rangle )$.

Analogously, we can construct $M^{(i)}(|\psi \rangle )$ as
\begin{equation}
M^{(i)}(|\psi \rangle )=\left(
\begin{tabular}{cc}
$P^{(i)}(|\psi \rangle )$ & $T^{(i)}(|\psi \rangle )$ \\
$T^{(i)}(|\psi \rangle )$ & $Q^{(i)}(|\psi \rangle )$%
\end{tabular}%
\right) .  \label{relation-3}
\end{equation}
Note that $M^{(i)}(|\psi \rangle )$ can also be obtained from
$M(|\psi \rangle )$ by taking transpositions $(1,i)$ on qubits 1 and
$i$. Clearly, $M^{(i)}(|\psi \rangle )$ is a symmetric matrix and
$M^{(1)}(|\psi \rangle )=M(|\psi \rangle )$. The rank of the matrix
$M^{(i)}(|\psi \rangle ) $ in Eq. (\ref{relation-3}) is referred to
as the rank of the state $|\psi \rangle $ with respect to qubit $i$
and denoted as $rank^{(i)}(|\psi \rangle )$.

For example, for three qubits, we obtain $rank^{(i)}(|W\rangle )=1$
for
$i=1$%
, $2$, $3$, whereas for any odd $n$ qubits, we find that $rank^{(i)}(|GHZ%
\rangle )=2$ and $rank^{(i)}(|0\cdots 0\rangle )=0$ for $i=1,
\cdots, n$.

Next, we establish the invariance of the rank for any state of odd
$n$ qubits under SLOCC. Let $|\psi ^{\prime }\rangle$ be another 
odd $n$-qubit state with 
$|\psi ^{\prime }\rangle=\sum_{i=0}^{2^{n}-1}b_{i}|i\rangle $. 
Recall that if two states
$|\psi\rangle $ and $|\psi ^{\prime }$ are SLOCC equivalent, then
there exist invertible local operators ${\mathcal{A}}_{1}$,
${\mathcal{A}}_{2}, \cdots$ and ${\mathcal{A}}_{n}$ ($\det
(A_i)\not=0$) such that \cite{Dur}
\begin{equation}
|\psi \rangle =\underbrace{{\mathcal{A}}_{1}\otimes
{\mathcal{A}}_{2}\otimes \cdots\otimes {\mathcal{A}}_{n}}_{n}|\psi
^{\prime }\rangle .  \label{odd-1}
\end{equation}%
Then, we assert that if $|\psi \rangle $ and $|\psi ^{\prime
}\rangle $ are SLOCC equivalent then the following SLOCC matrix equation
holds (see Appendix A for the proof):
\begin{equation}
M^{(i)}(|\psi \rangle )={\mathcal{A}}_{i}M^{(i)}(|\psi ^{\prime }\rangle )%
{\mathcal{A}}_{i}^{T}\det ({\mathcal{A}}_{1})\cdots \det
({\mathcal{A}}_{i-1})\det ({\mathcal{A}}_{i+1})\cdots \det
({\mathcal{A}}_{n}),  \label{relation-4}
\end{equation}%
where $M^{(i)}(|\psi ^{\prime }\rangle )$ is obtained from
$M^{(i)}(|\psi \rangle )$ by replacing $|\psi \rangle $ by $|\psi
^{\prime }\rangle $.

It follows from Eq. (\ref{relation-4}) that the rank of the matrix
$M^{(i)}(|\psi \rangle )$ in Eq. (\ref{relation-3}) is invariant
under SLOCC, thereby revealing that the rank of the state $|\psi
\rangle $ with respect to
qubit $%
i$ is an inherent property. Then the following result holds:
if two states are SLOCC equivalent, then they have
the same rank with respect to the same qubit $i$.
It should be noted that the converse does not hold, 
i.e., two states with the same rank with respect to
the same qubit are not necessarily equivalent.

To exemplify, we consider the $n$-qubit symmetric Dicke states
$|\ell,n\rangle $ with $\ell$ excitations, $1\leq \ell\leq (n-1)$
\cite{Stockton}: 
\begin{equation}
|\ell,n\rangle =\left( _{\ell}^{n}\right) ^{-1/2}\sum\limits_k
P_{k}|1_{1},1_{2},\cdots, 1_{\ell},0_{\ell+1},\cdots, 0_{n}\rangle,
\label{odd-1}
\end{equation}%
where $\{P_{k}\}$\ is the set of all distinct permutations of the spins. 
For any odd $n\geq 3$
qubits, a straightforward calculation yields
$rank^{(i)}(|(n-1)/2,n\rangle )=1$, $i=1,\cdots ,n$. For any odd
$n\geq 5$ qubits, $rank^{(i)}(|\ell,n\rangle )=0$ (note that
$rank^{(i)}(|W\rangle)=0$ as well, since $n$-qubit $|W\rangle$ state
is identical with $|1,n\rangle$) for $1\leq \ell<(n-1)/2$ and $i=1,\cdots
,n$. Since the Dicke states $|\ell,n\rangle $ and $|(n-\ell),n\rangle $
are SLOCC equivalent, the rank for any Dicke state can be determined.

Now consider pure biseparable states, i.e., those that are separable
under some bipartition.
By virtue of Theorem 3.4 of \cite{LDF-JMP}, we arrive at a necessary
condition for a pure state to be biseparable:
if $|\psi \rangle$ is a pure biseparable state of
odd $n$ qubits, then $rank^{(i)}(|\psi \rangle )=0$ or $1$ for some
$i$ with $1\leq i\leq n$.

In view of the above condition and the fact that a pure state of $n$ qubits is
genuinely entangled if it is not biseparable, we obtain the
following sufficient condition for a pure state to be genuinely
entangled:
for any pure state $|\psi \rangle $ of odd $n$
qubits, if $rank^{(i)}(|\psi \rangle )=2$ for any $1\leq i\leq n$,
then $|\psi \rangle$ is genuinely entangled.

\textit{Remark.} If we take the absolute value of the determinant
of $M^{(i)}(|\psi\rangle)$ given in Eq. (\ref{relation-3}), then we
obtain the $n$-tangle with respect to qubit $i$ of odd $n$ qubits
$\tau_{12\cdots n}^{(i)}$ given in \cite{LDFoddntangle} (up to a
constant factor). In particular, when $n=3$, $|\det
M(|\psi\rangle)|$ is, up to a constant factor, equal to the 3-tangle
\cite{Coffman} (we refer the reader to \cite{LDF07a} for more
details). Further, taking the determinants of both sides of Eq.
(\ref{relation-4}) yields
\begin{equation}
\det M^{(i)}(|\psi \rangle )=\det M^{(i)}(|\psi ^{\prime }\rangle )[\det (%
{\mathcal{A}}_{1})\cdots \det ({\mathcal{A}}_{n})]^{2}.
\label{relation-5}
\end{equation}
Note that for $i=1$, we recover Eq. (2.16) of \cite{LDF07a}. It
follows from Eq. (\ref{relation-5}) that if one of $\det
M^{(i)}(|\psi \rangle)$ and $\det M^{(i)}(|\psi ^{\prime }\rangle)$
vanishes while the other does not, then the state $|\psi\rangle$ is
not equivalent to $|\psi^{\prime} \rangle$ under SLOCC. Clearly, the
SLOCC invariance of the rank of $M^{(i)}(|\psi\rangle )$ is stronger
than the invariance of the determinant.

\section{SLOCC classification of odd $n$ qubits}

\subsection{Three families based on the rank with respect to qubit $i$}
\noindent The rank with respect to qubit $i$ permits the
partitioning of the space of the pure states of odd $n$ qubits into
the following three families: $F_{r_{i}}^{(i)}=\{|\psi \rangle
:rank^{(i)}(|\psi \rangle )=r_{i}\}$, $r_{i}\in \{0,1,2\}$. For
example, the rank with respect to qubit $1$ divides the space of the
pure states of odd $n$ qubits into three families:
$F_{0}^{(1)}=\{|\psi \rangle :rank^{(1)}(|\psi \rangle )=0\}$,
$F_{1}^{(1)}=\{|\psi \rangle :rank^{(1)}(|\psi \rangle )=1\}$, and
$F_{2}^{(1)}=\{|\psi \rangle :rank^{(1)}(|\psi \rangle )=2\}$.

It is not hard to see that two states belong to the same family if
and only if they have the same rank with respect to the same qubit.
Accordingly, if two states are SLOCC equivalent then they belong to
the same family $F_{r_{i}}^{(i)}$. However, the converse
does not hold, i.e., the states in the same family may be
inequivalent under SLOCC. It is further noted that the
aforementioned three SLOCC families $F_{0}^{(i)}$, $F_{1}^{(i)}$ and
$F_{2}^{(i)}$ form a complete partition of the space of odd $n$
qubits. That is, any state of odd $n$ qubits belongs to one and only
one of the above three families.

We exemplify the result for the six SLOCC equivalent classes for
three qubits: $|GHZ\rangle $, $|W\rangle $, $A$-$BC$, $B$-$AC$,
$C$-$AB$ and $A$-$B$-$C$ \cite{Dur}. The rank with respect to qubit
$i$ permits the partitioning of the space of three qubits into three
families $F_{0}^{(i)}$, $F_{1}^{(i)}$ and $F_{2}^{(i)}$, as
illustrated in Table \ref{table1}.

\begin{table}[h]
\caption{The three partitions for three qubits}
\centerline{\footnotesize
\begin{tabular}{lll}\\
\hline\hline
qubit $i$ & family & SLOCC classes\\
\hline
& $F_2^{(1)}$ & $|GHZ\rangle$ \\
$i=1$ & $F_1^{(1)}$ & $|W\rangle$, $A-BC$ \\
& $F_0^{(1)}$ & $A-B-C$, $B-AC$, $C-AB$ \\
\hline
& $F_2^{(2)}$ & $|GHZ\rangle$ \\
$i=2$ & $F_1^{(2)}$ & $|W\rangle$, $B-AC$ \\
& $F_0^{(2)}$ & $A-B-C$, $A-BC$, $C-AB$ \\
\hline & $F_2^{(3)}$ & $|GHZ\rangle$ \\
$i=3$ & $F_1^{(3)}$ & $|W\rangle$, $C-AB$ \\
& $F_0^{(3)}$ & $A-B-C$, $A-BC$, $B-AC$ \\
\hline\\
\end{tabular}}
\label{table1}
\end{table}

We also revisit the examples in the last section. Clearly, for any
odd $n\geq 5$ qubits, $|GHZ\rangle $ belongs to family
$F_{2}^{(i)}$, the Dicke state $|(n-1)/2,n\rangle $ belongs to
family $F_{1}^{(i)}$, whereas all the full separable states and all
the Dicke states $|\ell,n\rangle $ (including $n$-qubit $|W\rangle $ state) 
for $1\leq \ell<(n-1)/2$, belong to family $F_{0}^{(i)}$, $i=1, \cdots, n$.

\subsection{Nine families based on the ranks with respect to qubits $1$ and
$2$} \noindent As discussed in the previous section, 
the rank with respect to qubit $1$
divides the space of odd $n$ qubits into three families
$F_{0}^{(1)}$, $F_{1}^{(1)}$ and $F_{2}^{(1)}$. For odd $n\geq 5$
qubits, based on the rank with respect to qubit 2 each family
$F_{r_{1}}^{(1)}$, $r_{1}\in \{0$, $1$, $2\}$, can be further
divided into three different families:
$F_{r_{1},r_{2}}^{(1,2)}=F_{r_{1}}^{(1)}\cap F_{r_{2}}^{(2)}$,
$r_{2}\in \{0$, $1$, $2\}$. Here, each family
$F_{r_{1},r_{2}}^{(1,2)}$ is the intersection of the families
$F_{r_{1}}^{(1)}$ and $F_{r_{2}}^{(2)}$. More specifically, the
family $F_{2}^{(1)}$ is divided into three families
$F_{2,0}^{(1,2)}$, $F_{2,1}^{(1,2)}$ and $F_{2,2}^{(1,2)}$, the
family $F_{1}^{(1)}$ into three families $F_{1,0}^{(1,2)}$,
$F_{1,1}^{(1,2)}$ and $F_{1,2}^{(1,2)}$, and the family
$F_{0}^{(1)}$ into three families $F_{0,0}^{(1,2)}$,
$F_{0,1}^{(1,2)}$ and $F_{0,2}^{(1,2)}$. For odd $n\geq 5$ qubits,
we list the representative states of the families
$F_{r_{1},r_{2}}^{(1,2)}$ in Table \ref{table2}.

\begin{table}[h]
\caption{The nine families for odd $n\geq 5$ qubits based on the
ranks with respect to qubits 1, 2}
\centerline{\footnotesize
\begin{tabular}{ll}\\
\hline\hline
family & representative state\\
\hline
$F_{2,2}^{(1,2)}$ & $|GHZ\rangle$ \\
$F_{2,1}^{(1,2)}$ & $\frac{1}{\sqrt{6}}[(|0\cdots 0\rangle+|1\cdots
1\rangle) +(|010\cdots 010\rangle+|101\cdots 101\rangle)+(|0\cdots
0110\rangle
-|101\cdots 10001 \rangle)]$ \\
$F_{2,0}^{(1,2)}$ & $\frac{1}{2}[(|0\cdots 0\rangle+|1\cdots
1\rangle)
+(|010\cdots 010\rangle+|101\cdots 101\rangle)]$ \\
$F_{1,2}^{(1,2)}$ & $\frac{1}{\sqrt{5}}[(|001\cdots 1\rangle
-|010\cdots 0\rangle)
+(|0110\cdots 0\rangle+|10\cdots 0\rangle)+|1101\cdots 1\rangle]$ \\
$F_{1,1}^{(1,2)}$ & $|(n-1)/2,n\rangle$ \\
$F_{1,0}^{(1,2)}$ & $\frac{1}{\sqrt{2}} (|0\cdots 0\rangle+|01\cdots
1\rangle)$ \\
$F_{0,2}^{(1,2)}$ & $\frac{1}{2}(|0\cdots 0\rangle +|1\cdots
1\rangle
+|010\cdots 01\rangle -|101\cdots 10\rangle)$ \\
$F_{0,1}^{(1,2)}$ & $\frac{1}{\sqrt{2}}(|0\cdots 0\rangle
+|101\cdots
1\rangle )$ \\
$F_{0,0}^{(1,2)}$ & $|0\cdots 0\rangle $ \\
\hline\\
\end{tabular}}
\label{table2}
\end{table}

Consequently, the ranks with respect to qubits 1 and 2 divide the
space of odd $n\geq 5$ qubits into nine different families. Note
furthermore that the nine SLOCC families form a complete partition
of the space of odd $n\geq 5$ qubits. That is, any state of odd $n$
qubits belongs to one and only one of the nine families.

Continuing with the example for three qubits, we see that the six
SLOCC equivalence classes are divided into five families based on
the ranks with respect to qubits 1 and 2, see Table \ref{table3}.

\begin{table}[h]
\caption{Partition for three qubits based on the ranks with respect
to qubits 1, 2} \centerline{\footnotesize
\begin{tabular}{ll}\\
\hline\hline
family & SLOCC equivalent class \\
\hline
$F_{2,2}^{(1,2)}$ & $|GHZ\rangle$ \\
$F_{1,1}^{(1,2)}$ & $|W\rangle$ \\
$F_{1,0}^{(1,2)}$ & $A-BC$ \\
$F_{0,1}^{(1,2)}$ & $B-AC$ \\
$F_{0,0}^{(1,2)}$ & $A-B-C$, $C-AB$ \\
\hline\\
\end{tabular}}
\label{table3}
\end{table}

\subsection{$3^{n}$ families based on the ranks with respect to qubits $1,
\cdots, n$} \noindent 
Now, assume that the ranks with respect to qubits $1,
\cdots, (\ell-1)$ permit the partitioning of the space of odd $n\geq 5$
qubits into $3^{(\ell-1)}$ families: $F_{r_{1},r_{2},\cdots
,r_{(\ell-1)}}^{(1,2,\cdots ,(\ell-1))}$, $r_{1},r_{2},\cdots
,r_{(\ell-1)}\in \{0,1,2\}$. Then, each family $F_{r_{1},r_{2},\cdots
,r_{(\ell-1)}}^{(1,2,\cdots ,(\ell-1))}$ can be further divided into three
families: $F_{r_{1},r_{2},\cdots ,r_{\ell}}^{(1,2,\cdots,\ell)}
=F_{r_{1},r_{2},\cdots ,r_{(\ell-1)}}^{(1,2,\cdots ,(\ell-1))}\cap
F_{r_{\ell}}^{(\ell)}$, $r_{\ell}\in \{0,1,2\}$ based on the rank with
respect to qubit $\ell$. Clearly, each family $F_{r_{1},r_{2},\cdots
,r_{\ell}}^{(1,2,\cdots,\ell)}$ is associated with the sequence
$\{r_{1},\cdots ,r_{\ell}\}$, $r_{1},\cdots ,r_{\ell}\in \{0,1,2\}$, and
different sequences correspond to different families. Consequently,
in total there are $3^{\ell}$ SLOCC families based on the ranks with
respect to qubits $1, \cdots, \ell$. In particular, there are $3^{n}$
SLOCC families based on the ranks with respect to qubits $1, \cdots,
n$. It should be noted that at least one family contains an infinite
number of SLOCC classes.

It is readily seen that $n$-qubit $|GHZ\rangle $ state belongs to
family $F_{2,\cdots ,2}^{(1,\cdots ,n)}$, the Dicke states
$|(n-1)/2,n\rangle $ and $|(n+1)/2,n\rangle $ belong to family
$F_{1,\cdots ,1}^{(1,\cdots ,n)}$, whereas all the full separable
states and all the Dicke states $|\ell,n\rangle $ (including $n$-qubit 
$|W\rangle $ state), 
with $1\leq \ell<(n-1)/2$ and $n\ge 5$, 
belong to family $F_{0,\cdots ,0}^{(1,\cdots
,n)}$. It is worth pointing out that all the states in the family
$F_{2,\cdots ,2}^{(1,\cdots ,n)}$ are genuinely entangled as discussed
in section 2.

For any state $|\psi \rangle$ of odd $n$ qubits, by computing
$rank^{(i)}(|\psi \rangle )$, $i=1, \cdots, \ell$ $(\leq n)$, we can
determine which family the state $|\psi \rangle $ belongs to. It is
plain to see that two states belong to the same family if and only
if they have the same ranks with respect to qubits $1, \cdots, \ell$
$(\leq n)$. Thus, if two states are SLOCC equivalent then they
belong to the same family.

Consider once again the example for three qubits. A straightforward
calculation demonstrates that the six SLOCC equivalence classes of
three qubits are divided into six families based on the ranks with
respect to qubits 1, 2 and 3, i.e., each family is just a single
SLOCC class, see Table \ref{table4}.

\begin{table}[h]
\caption{Partition for three qubits based on the ranks with respect
to qubits 1, 2, 3} \centerline{\footnotesize
\begin{tabular}{ll}\\
\hline\hline
family & SLOCC equivalent class \\
\hline
$F_{2,2,2}^{(1,2,3)}$ & $|GHZ\rangle$ \\
$F_{1,1,1}^{(1,2,3)}$ & $|W\rangle$ \\
$F_{1,0,0}^{(1,2,3)}$ & $A-BC$ \\
$F_{0,1,0}^{(1,2,3)}$ & $B-AC$ \\
$F_{0,0,1}^{(1,2,3)}$ & $C-AB$ \\
$F_{0,0,0}^{(1,2,3)}$ & $A-B-C$ \\
\hline\\
\end{tabular}}
\label{table4}
\end{table}

\section{Discussion}
\noindent In \cite{Osterloh06}, the ``filter"
approach was used to separate SLOCC orbits and it was shown 
that the following four five-qubit states
\begin{eqnarray}
|\Phi _{1}\rangle &=&\frac{1}{\sqrt{2}}(|11111\rangle+|00000\rangle), \\
|\Phi _{2}\rangle &=&\frac{1}{2}(|11111\rangle +|11100\rangle
+|00010\rangle +|00001\rangle), \\
|\Phi _{3}\rangle &=&\frac{1}{\sqrt{6}}(
\sqrt{2}|11111\rangle +|11000\rangle +|00100\rangle +|00010\rangle
+|00001\rangle ), \\
|\Phi _{4}\rangle &=&\frac{1}{2\sqrt{2}}(\sqrt{3}
|11111\rangle +|10000\rangle +|01000\rangle +|00100\rangle
+|00010\rangle +|00001\rangle )
\end{eqnarray} 
are in different orbits. 

We now classify the above four states using our framework. Based on
the
ranks with respect to qubits 1, 2 and 3, $|\Phi _{1}\rangle$ belongs to $%
F_{2,2,2}^{(1,2,3)}$, $|\Phi _{2}\rangle $ belongs to
$F_{0,0,0}^{(1,2,3)}$, $|\Phi _{3}\rangle$ belongs to
$F_{0,0,1}^{(1,2,3)}$, and $|\Phi _{4}\rangle$ belongs to
$F_{1,1,1}^{(1,2,3)}$, in agreement with \cite{Osterloh06} that the
four states are in different orbits.

Also note that the space of five qubits is divided into nine
different families based on the ranks with respect to qubits 1 and
2. We list the representatives of the nine families in Table
\ref{table5}.

\begin{table}[h]
\caption{The nine families for five qubits based on the ranks with
respect to qubits 1, 2} \centerline{\footnotesize
\begin{tabular}{ll}\\
\hline\hline
family & representative state \\
\hline
$F_{2,2}^{(1,2)}$ & $|GHZ\rangle$ \\
$F_{2,1}^{(1,2)}$ & $\frac{1}{\sqrt{6}}(|00000\rangle+|11111\rangle
+|01010\rangle+|10101\rangle+|00110\rangle-|10001\rangle)$ \\
$F_{2,0}^{(1,2)}$ & $\frac{1}{2}(|00000\rangle+|11111\rangle
+|01010\rangle+|10101\rangle)$ \\
$F_{1,2}^{(1,2)}$ & $\frac{1}{\sqrt{5}}(|00111\rangle-|01000\rangle
+|01100\rangle+|10000\rangle+|11011\rangle)$ \\
$F_{1,1}^{(1,2)}$ & $|2,5\rangle$ \\
$F_{1,0}^{(1,2)}$ & $\frac{1}{\sqrt{2}} (|00000\rangle +|01111\rangle )$ \\
$F_{0,2}^{(1,2)}$ & $\frac{1}{2}(|00000\rangle +|11111\rangle
+|01001\rangle
-|10110\rangle) $ \\
$F_{0,1}^{(1,2)}$ & $\frac{1}{\sqrt{2}}(|00000\rangle +|10111\rangle ) $ \\
$F_{0,0}^{(1,2)}$ & $|00000\rangle $ \\
\hline\\
\end{tabular}}
\label{table5}
\end{table}

\section{Conclusion}
\noindent In this paper, we have introduced the rank with respect to
qubit $i$ for any state of odd $n$ qubits and established its
invariance under SLOCC. That is, if two states are SLOCC equivalent
then they have the same ranks with respect to qubits $1, \cdots, n$.
The ranks with respect to qubits $1, \cdots, n$ permit the
partitioning of the space of odd $n\geq 5$ qubits into $3^{n}$
inequivalent families. It is straightforward to know that two states
belong to the same family if and only if they have the same ranks
with respect to qubits $1, \cdots, n$. In other words, all the
states of a family have the same ranks with respect to qubits $1,
\cdots, n$. As a consequence, if two states are SLOCC equivalent
then they belong to the same family. Furthermore, each family
corresponds to the sequence $\{r_{1},\cdots ,r_{n}\}$, $r_{i}\in
\{0,1,2\}$, and different families correspond to different
sequences. In terms of the ranks, we have given a necessary
condition for a pure state to be biseparable and a sufficient
condition for a pure state to be genuinely entangled. The
classification based on the ranks of states may possess more
physical meaning. As a final note, we would like to mention that
the SLOCC invariance of the rank for odd $n$ qubits
does not hold for even $n$ qubits.

\bigskip
{\bf Acknowledgement.} \noindent We would like to thank
Thierry Bastin for helpful discussions and the reviewers for the
useful comments. This work was supported by NSFC (Grant No.
10875061) and Tsinghua National Laboratory for Information Science
and Technology.

\section*{Appendix}

\noindent We here give the proof of Eq. (\protect\ref{relation-4}).
We distinguish two cases: $i=1$ and $2\leq i\leq n$.

\textit{Case 1.} $i=1$.

In this case, Eq. (\ref{relation-4}) becomes
\begin{equation}
M(|\psi \rangle )={\mathcal{A}}_{1}M(|\psi ^{\prime }\rangle
){\mathcal{A}} _{1}^{T}\det ({\mathcal{A}}_{2})\cdots \det
({\mathcal{A}}_{n}). \label{relation-2}
\end{equation}
Let $|\psi \rangle $ and $|\psi ^{\prime }\rangle $ be related by
Eq. (\ref{odd-1}), and ${\mathcal{A}}_{1}=\left(
\begin{tabular}{ll}
$\alpha _{1}$ & $\alpha _{2}$ \\
$\alpha _{3}$ & $\alpha _{4}$%
\end{tabular}%
\right) $. It is easy to verify that Eq. (\ref{relation-2}) holds if
and only if the following three SLOCC equations hold together:
\begin{eqnarray}
T(|\psi \rangle ) &=&[P(|\psi ^{\prime }\rangle )\alpha _{1}\alpha
_{3}+T(|\psi ^{\prime }\rangle )(\alpha _{2}\alpha _{3}+\alpha
_{1}\alpha
_{4})+Q(|\psi ^{\prime }\rangle )\alpha _{2}\alpha _{4}]  \nonumber \\
&&\times \det ({\mathcal{A}}_{2})\cdots \det ({\mathcal{A}}_{n}),
\label{SLOCC-2} \\
P(|\psi \rangle ) &=&[P(|\psi ^{\prime }\rangle )\alpha
_{1}^{2}+2T(|\psi ^{\prime }\rangle )\alpha _{1}\alpha _{2}+Q(|\psi
^{\prime }\rangle )\alpha
_{2}^{2}]  \nonumber \\
&&\times \det ({\mathcal{A}}_{2})\cdots \det ({\mathcal{A}}_{n}),
\label{SLOCC-3} \\
Q(|\psi \rangle ) &=&[P(|\psi ^{\prime }\rangle )\alpha
_{3}^{2}+2T(|\psi ^{\prime }\rangle )\alpha _{3}\alpha _{4}+Q(|\psi
^{\prime }\rangle )\alpha
_{4}^{2}]  \nonumber \\
&&\times \det ({\mathcal{A}}_{2})\cdots \det ({\mathcal{A}}_{n}).
\label{SLOCC-4}
\end{eqnarray}

Notice that ${\mathcal{A}}_{1}\otimes {\mathcal{A}}_{2}\otimes
\cdots \otimes {\mathcal{A}}_{n}$ can be written as
$({\mathcal{A}}_{1}\otimes {\mathcal{I}} _{2}\otimes \cdots \otimes
{\mathcal{I}}_{n})\circ ({\mathcal{I}}_{1}\otimes
{\mathcal{A}}_{2}\otimes {\mathcal{I}}_{3}\otimes \cdots \otimes
{\mathcal{I}} _{n})\circ \cdots \circ ({\mathcal{I}}_{1}\otimes
\cdots \otimes {\mathcal{I}} _{n-1}\otimes {\mathcal{A}}_{n})$, then
Eqs. (\ref{SLOCC-2}), (\ref{SLOCC-3}) and (\ref{SLOCC-4}) follow
immediately from the two lemmas below.

\textit{Lemma 1}. For odd $n$ qubits, if $|\psi \rangle $ and $|\psi
^{\prime }\rangle $ are related by
\begin{equation}
|\psi \rangle =\underbrace{{\mathcal{A}}_{1}\otimes
{\mathcal{I}}_{2}\otimes \cdots \otimes {\mathcal{I}}_{n}}_{n}|\psi
^{\prime }\rangle , \label{SLOCC-5}
\end{equation}%
then%
\begin{eqnarray}
T(|\psi \rangle ) &=&P(|\psi ^{\prime }\rangle )\alpha _{1}\alpha
_{3}+T(|\psi ^{\prime }\rangle )(\alpha _{2}\alpha _{3}+\alpha
_{1}\alpha
_{4})+Q(|\psi ^{\prime }\rangle )\alpha _{2}\alpha _{4},  \label{SLOCC-6} \\
P(|\psi \rangle ) &=&P(|\psi ^{\prime }\rangle )\alpha
_{1}^{2}+2T(|\psi ^{\prime }\rangle )\alpha _{1}\alpha _{2}+Q(|\psi
^{\prime }\rangle )\alpha
_{2}^{2},  \label{SLOCC-7} \\
Q(|\psi \rangle ) &=&P(|\psi ^{\prime }\rangle )\alpha
_{3}^{2}+2T(|\psi ^{\prime }\rangle )\alpha _{3}\alpha _{4}+Q(|\psi
^{\prime }\rangle )\alpha _{4}^{2}.  \label{SLOCC-8}
\end{eqnarray}

\textit{Proof.} We only prove Eq. (\ref{SLOCC-6}). The proofs for
Eqs. (\ref{SLOCC-7}) and (\ref{SLOCC-8}) are analogous. By Eq.
(\ref{SLOCC-5}), we obtain
\begin{equation}
a_{i}=\alpha _{1}b_{i}+\alpha _{2}b_{2^{n-1}+i},\quad
a_{2^{n-1}+i}=\alpha _{3}b_{i}+\alpha _{4}b_{2^{n-1}+i},  \label{D0}
\end{equation}
for $0\leq i\leq 2^{n-1}-1$. By substituting Eq. (\ref{D0}) into Eq.
(\ref{T-1}%
), we obtain
\begin{equation}
T(|\psi \rangle )=\sum_{i=0}^{2^{n-1}-1}(-1)^{N(i)}(\alpha
_{1}b_{i}+\alpha _{2}b_{2^{n-1}+i})(\alpha
_{3}b_{2^{n-1}-i-1}+\alpha _{4}b_{2^{n}-i-1}). \label{T-3}
\end{equation}%
Note that $T(|\psi ^{\prime }\rangle )$, $P(|\psi ^{\prime }\rangle
)$, and $Q(|\psi ^{\prime }\rangle )$ can be rewritten as:
\begin{eqnarray}
T(|\psi ^{\prime }\rangle )
&=&\sum_{i=0}^{2^{n-1}-1}(-1)^{N(i)}b_{2^{n-1}+i}b_{2^{n-1}-i-1},
\label{T-5} \\
P(|\psi ^{\prime }\rangle )
&=&\sum_{i=0}^{2^{n-1}-1}(-1)^{N(i)}b_{i}b_{2^{n-1}-i-1},  \label{P-4} \\
Q(|\psi ^{\prime }\rangle )
&=&\sum_{i=0}^{2^{n-1}-1}(-1)^{N(i)}b_{2^{n-1}+i}b_{2^{n}-i-1}.
\label{Q-4}
\end{eqnarray}

Expanding Eq. (\ref{T-3}) and using Eqs. (\ref{T-5}), (\ref{P-4}) and (\ref%
{Q-4}) yield Eq. (\ref{SLOCC-6}).

\textit{Lemma 2.} For odd $n$ qubits, if $|\psi \rangle $ and $|\psi
^{\prime }\rangle $ are related by
\begin{equation}
|\psi \rangle =\underbrace{{\mathcal{I}}_{1}\otimes \cdots \otimes
{\mathcal{I}} _{k-1}\otimes {\mathcal{A}}_{k}\otimes
{\mathcal{I}}_{k+1}\otimes \cdots \otimes
{\mathcal{I}}_{n}}_{n}|\psi ^{\prime }\rangle ,  \label{SLOCC-9}
\end{equation}%
then%
\begin{eqnarray}
T(|\psi \rangle ) &=&T(|\psi ^{\prime }\rangle )\det
({\mathcal{A}}_{k}),
\label{eq-1} \\
P(|\psi \rangle ) &=&P(|\psi ^{\prime }\rangle )\det
({\mathcal{A}}_{k}),
\label{eq-2} \\
Q(|\psi \rangle ) &=&Q(|\psi ^{\prime }\rangle )\det
({\mathcal{A}}_{k}), \label{eq-3}
\end{eqnarray}%
for $2\leq k\leq n$.

\textit{Proof. }We only prove Eq. (\ref{eq-1}). The proofs for Eqs. (\ref%
{eq-2}) and (\ref{eq-3}) can be given analogously. It is sufficient
to consider $k=2$. Let ${\mathcal{A}}_{2}=\left(
\begin{tabular}{cc}
$\beta _{1}$ & $\beta _{2}$ \\
$\beta _{3}$ & $\beta _{4}$%
\end{tabular}%
\right) $. Then, by Eq. (\ref{SLOCC-9}), we obtain
\begin{eqnarray}
a_{i} &=&\beta _{1}b_{i}+\beta _{2}b_{2^{n-2}+i},  \label{coef-1} \\
a_{2^{n-2}+i} &=&\beta _{3}b_{i}+\beta _{4}b_{2^{n-2}+i},  \label{coef-2} \\
a_{2^{n-1}+i} &=&\beta _{1}b_{2^{n-1}+i}+\beta
_{2}b_{2^{n-1}+2^{n-2}+i},
\label{coef-3} \\
a_{2^{n-1}+2^{n-2}+i} &=&\beta _{3}b_{2^{n-1}+i}+\beta
_{4}b_{2^{n-1}+2^{n-2}+i},  \label{coef-4}
\end{eqnarray}%
for $0\leq i\leq 2^{n-2}-1$. We may rewrite $T(|\psi \rangle )$ in
Eq.
(\ref{T-1}%
) as
\begin{eqnarray}
T(|\psi \rangle )
&=&\sum_{i=0}^{2^{n-2}-1}(-1)^{N(i)}a_{i}a_{2^{n}-i-1}
\nonumber \\
&&-\sum_{i=0}^{2^{n-2}-1}(-1)^{N(i)}a_{2^{n-2}+i}a_{2^{n-1}+2^{n-2}-i-1}
\nonumber \\
&&-\sum_{i=0}^{2^{n-2}-1}(-1)^{N(i)}a_{2^{n-1}+i}a_{2^{n-1}-i-1}
\nonumber
\\
&&+\sum_{i=0}^{2^{n-2}-1}(-1)^{N(i)}a_{2^{n-1}+2^{n-2}+i}a_{2^{n-2}-i-1}.
\label{inv-1}
\end{eqnarray}
Substituting Eqs. (\ref{coef-1}), (\ref{coef-2}), (\ref{coef-3}) and (\ref%
{coef-4}) into Eq. (\ref{inv-1}) yields the desired result Eq.
(\ref{eq-1}).

\textit{Case 2.} $2\leq i\leq n$.

We give a brief proof here. After a tedious but straightforward
calculation, the following identity holds:
\begin{equation}
(1,i)\circ ({\mathcal{A}}_{1}\otimes \cdots \otimes
{\mathcal{A}}_{n})\circ (1,i)=%
{\mathcal{A}}_{i}\otimes {\mathcal{A}}_{2}\otimes \cdots
{\mathcal{A}}_{i-1}\otimes {\mathcal{A}}_{1}\otimes
{\mathcal{A}}_{i+1}\otimes \cdots \otimes {\mathcal{A}}_{n}.
\label{tensor}
\end{equation}%
Letting $M^{(i)}=M\circ (1,i)$ and using Eq. (\ref{odd-1}), we have
\begin{eqnarray}
M^{(i)}(|\psi \rangle ) &=&M^{(i)}({\mathcal{A}}_{1}\otimes \cdots
\otimes
{\mathcal{A}}_{n}|\psi ^{\prime }\rangle )  \nonumber \\
&=&M\circ (1,i)\circ ({\mathcal{A}}_{1}\otimes \cdots \otimes
{\mathcal{A}} _{n})\circ (1,i)((1,i)|\psi ^{\prime }\rangle ).
\label{odd-2}
\end{eqnarray}%
By substituting Eq. (\ref{tensor}) into Eq. (\ref{odd-2}), then using Eq. (%
\ref{relation-2}), we obtain that
\begin{equation}
M^{(i)}(|\psi \rangle )={\mathcal{A}}_{i}M((1,i)|\psi ^{\prime }\rangle )%
{\mathcal{A}}_{i}^{T}\det ({\mathcal{A}}_{1})\cdots \det
({\mathcal{A}}_{i-1})\det ({\mathcal{A}}_{i+1})\cdots \det
({\mathcal{A}}_{n}),  \label{odd-4}
\end{equation}%
and then Eq. (\ref{relation-4}) follows immediately.

\end{document}